\documentclass[preprint,aps]{revtex4}
\usepackage{amsfonts}
\usepackage{graphicx}
\newcommand{\be}{\begin{equation}}
\newcommand{\ee}{\end{equation}}
\newcommand{\ba}{\begin{eqnarray}}
\newcommand{\ea}{\end{eqnarray}}

\newcommand{\f}{\frac}

\begin{document}
\title
{Formulation of general dynamical invariants and their unitary relations for time-dependent
three coupled quantum oscillators
 \vspace{0.3cm}}
\author{Jeong Ryeol Choi\footnote{E-mail: choiardor@hanmail.net } \vspace{0.3cm}}

\affiliation{Department of Nanoengineering, Kyonggi University, 
Yeongtong-gu, Suwon,
Gyeonggi-do 16227, Republic of Korea \vspace{0.7cm}}

\begin{abstract}
\indent
A general dynamical invariant operator for three coupled time-dependent oscillators is derived.
Although the obtained invariant operator satisfies the Liouville-von Neumann equation,
its mathematical formula is somewhat complicated due to arbitrariness of time variations of
parameters.
The parametric conditions required for formulating this invariant are definitely specified.
By using the unitary transformation method, the invariant operator is transformed
to the one that corresponds to three independent simple harmonic oscillators.
Inverse transformation of the well-known quantum solutions associated with such a simplified invariant
enables us to identify quantum solutions of the coupled original systems.
These solutions are exact since we do not use approximations not only in formulating
the invariant operator but in the unitary transformation as well.
The invariant operator and its eigenfunctions provided here can be used
to characterize quantum properties of the systems with various choices of the types of
time-dependent parameters.
\end{abstract}

\maketitle
\newpage

{\ \ \ } \\
{\bf 1. Introduction
\vspace{0.2cm}}
\\
A large part of modern quantum technologies are
based on the utilization of
entanglement between identical or different quantum devices,
such as qubits, micro cavities,
quantum dot transistors, nano resonators, and tunnel diodes.
Generation and control of entanglement with high precision
are requisite in order to
build large-scale quantum architectures in quantum information science \cite{pce,pce2,pce3}.
Hence it is important
to understand quantum entanglement in coupled devices from
a quantum-mechanical point of view.
Lots of quantum information devices are analyzed using
a model of coupled oscillators \cite{qde,qde2,qde3}.
As a next-generation technological resource along this line,
Boolean computations organized by means of interacting oscillators in
quantum circuits \cite{qde} are expected to
play a potential role in solving challenging computational problems,
such as prime factorizations of large numbers and NP-complete problems.
A model of coupled oscillators can also be applied in the research
of other scientific branches,
such as electromagnetic induced transparency \cite{ab,ab2}, periodicity
of solar activity \cite{saa}, locomotion gaits of bio-inspired robots \cite{lgi,aot}, and
coherence in coupled semiconductor lasers \cite{csl}.

To analyze coupled oscillatory systems accurately,
their exact quantum formalism established on the basis of
Hamiltonian diagonalization is necessary.
However, if we diagonalize the Hamiltonian directly, there may arise an additional term
in the Hamiltonian in addition to the diagonalization term.
For the details of such an additional term, see the second term of Eq. (48) in Ref. \cite{VERSION1} or
the second term of Eq. (3) in Ref. \cite{ada} for examples.
Because we do not know how to manage
such a term exactly in the diagonalization of the Hamiltonian,
it may be better to adopt an alternative method.

We introduce an invariant operator as such an alternative \cite{Lewis1,Lewis2}
and diagonalize it instead of directly diagonalizing the Hamiltonian.
Because an additional
term does not appear in the diagonalization of the invariant,
we only need to diagonalize the invariant itself in this case.
The diagonalization of a dynamical invariant for two coupled
time-dependent oscillators has already been carried out
by ours in Ref. \cite{tbo} considering such an advantage.
We extend it to three coupled time-dependent oscillators
and decouple the couplings in the oscillators in this work.
However, the mathematical manage of the operator may not be so easy in
this case because of the three coupling terms in the invariant
in addition to the time variations of parameters.

We will organize this work as follows.
In Sec. 2, we will establish an invariant operator for the Hamiltonian of time-dependent
three coupled oscillators.
Parametric conditions required
for such an invariant formulation will be found and specified.
By using the unitary transformation method,
we will transform the invariant into a simple form in Sec. 3, which is identical to the collection of
three Hamiltonians of the simple harmonic oscillators (SHOs).
We can regard the Hamiltonian itself as an invariant operator intrinsically
for the case of the SHO.
The invariant operator will be transformed in two steps.
At first, the invariant operator will be simplified by a preliminary transformation
using an appropriate unitary operator.
Through the next transformation, the invariant operator will be
finally reduced to that of three independent SHOs that are much simpler:
that is, the invariant operator will be diagonalized.
The eigenfunctions and eigenvalues of such a simplified (i.e., transformed)
invariant operator are well known from basic quantum mechanics.
Eventually, in Sec. 4, the eigenfunctions in the original systems will be identified
by inverse transformation of the ones in the transformed systems.
Some concluding remarks will be given in the last section.
\\
\\
{\bf 2. Formulation of the Invariant \vspace{0.2cm}} \\
We introduce the Hamiltonian of time-dependent three coupled oscillators as
\begin{eqnarray}
\hat{\mathcal H}(t) &=& \frac{1}{2}\sum_{j=1}^{3}\left[
\frac{\hat{p}_{j}^{2}}{m_{j}( t) }+b_j(t)(\hat{x}_j\hat{p}_j+\hat{p}_j\hat{x}_j)
+m_{j}( t) \omega _{j}^{2}( t)
\hat{x}_{j}^{2}\right] \nonumber \\
& &+ d_{12}( t)
\hat{x}_{1}\hat{x}_{2}+d_{13}( t) \hat{x}_{1}\hat{x}_{3}+ d_{23}( t) \hat{x}_{2}\hat{x}_{3} ,\   \label{1}
\end{eqnarray}%
where the parameters $m_{j}( t)$, $b_j(t)$, $\omega _{j}( t)$, and $d_{jk}( t)$ vary
over time, but in a fashion that they are differentiable with respect to time.
The coordinates are coupled via $d_{jk}(t)$ terms in this Hamiltonian as can be seen.
Additionally, this Hamiltonian involves $b_j(t)$ terms that are frequently appeared
in the mathematical treatment of damped oscillatory systems \cite{do2,tla,do1,do1-1}.

Before we start to formulate
a quantum invariant, let us briefly study the classical behavior of
oscillators based on their classical equations of motion which are
\begin{eqnarray}
\ddot{x}_{1}+\f{\dot{m}_{1}}{m_{1}} \dot{x}_{1}+
\tilde{\omega}_{1}^{2}x_{1}
+ \f{d_{12}}{m_1} x_{2}+\f{d_{13}}{m_1}x_{3} &=&0,  \label{2} \\
\ddot{x}_{2}+\f{\dot{m}_{2}}{m_{2}} \dot{x}_{2}
+\tilde{\omega}_{2}^{2}x_{2}
+ \f{d_{12}}{m_2} x_{1}+\f{d_{23}}{m_2}x_{3} &=&0,  \label{3} \\
\ddot{x}_{3}+\f{\dot{m}_{3}}{m_{3}} \dot{x}_{3}
+\tilde{\omega}_{3}^{2}x_{3}
+ \f{d_{13}}{m_3} x_{1}+\f{d_{23}}{m_3}x_{2} &=&0.  \label{4}
\end{eqnarray}%
Here, $\tilde{\omega}_{j}$ are modified angular frequencies of the form
\be
\tilde{\omega}_{j} = \bigg(\omega_{j}^{2}-b_j^2-\dot{b}_j
-b_j \f{\dot{m}_j}{m_j} \bigg)^{1/2}.  \label{5}
\ee
Equation (\ref{2}) [Eq. (\ref{3}), Eq. (\ref{4})] is not represented in terms of the canonical
variable $x_1$ [$x_2$, $x_3$] only, owing to the fact that
the three variables are coupled unless $d_{jk} = 0$.
Thereby the motion of each oscillator is affected by that of
other oscillators through couplings.
For instance, Eq. (\ref{2}) reveals
that the effect of oscillator 2 (oscillator 3) on oscillator 1
is large when $d_{12}$ ($d_{13}$) is great; but it is relatively small when $m_1$ is large.
The classical motion of other oscillators can also be interpreted in the same way
through the use of Eqs. (\ref{3}) and (\ref{4}).

Let us now formulate a quadratic invariant for the time-dependent coupled oscillators, which is useful in
developing quantum theory of the systems.
We assume that the invariant operator is represented in the form
\ba
\hat{\mathcal I} (t) &=& \f{1}{2}\sum_{j=1}^{3}\Big[
\alpha_{j}(t)  \hat{p}_{j}^{2}+\beta_{j}(t)  \left(
\hat{x}_{j}\hat{p}_{j}+\hat{p}_{j}\hat{x}_{j}\right)  +\gamma_{j}(t)  \hat{x}_{j}^{2}\Big] \nonumber \\
& &+\delta_{12}(t)  \hat{x}_{1}\hat{x}_{2}+\delta_{13}(t)  \hat{x}_{1}\hat{x}_{3}+\delta_{23}(t)  \hat{x}_{2}\hat{x}_{3},\label{6}%
\ea
where $\alpha_{j}(t)$, $\beta_{j}(t)$, $\gamma_{j}(t)$, and $\delta_{jk}(t)$ are time-dependent coefficients
that will be derived now.
We take the dimension of $\hat{\mathcal I}(t)$ as energy in this case as in 
the two coupled oscillators managed in Ref. \cite{tbo}.
By using the Liouville-von Neumann equation,
\begin{equation}
\frac{d\hat{\mathcal I}}{dt}=\frac{\partial \hat{\mathcal I}}{\partial t}+ \f{1}{i\hbar}
[\hat{\mathcal I}, \hat{\mathcal H}]=0,
  \label{7}
\end{equation}%
we can confirm that the coefficients should satisfy the equations
\begin{equation}
\dot{\alpha}_{j}(t) =2b_j(t)\alpha_j(t)-\frac{2\beta_{j}( t) }{m_{j}(t)},
\label{8}
\end{equation}%
\begin{equation}
\dot{\beta}_{j}( t) =m_{j}( t)\alpha_{j}( t) \omega_{j}^{2}(t)  -\frac{\gamma_{j}(t)}{m_{j}(t)},
\label{9}
\end{equation}%
\begin{equation}
\dot{\gamma}_{j}(t)=-2b_j(t)\gamma_j(t)+2m_{j}( t) \beta_{j}(t) \omega _{j}^{2}(t) ,  \label{10}
\end{equation}%
\begin{eqnarray}
\dot{\delta}_{12}(t) &=&-\delta_{12}(t)[ b_{1}(t) +b_{2}(t) ]+d_{12}(t)[ \beta_{1}(t) +\beta_{2}(t) ] ,  \label{11} \\
\dot{\delta}_{13}(t) &=&-\delta_{13}(t)[ b_{1}(t) +b_{3}(t) ]+d_{13}(t)[ \beta_{1}( t) +\beta_{3}(t) ] ,  \label{12} \\
\dot{\delta}_{23}(t) &=&-\delta_{23}(t)[ b_{2}(t) +b_{3}(t) ]+d_{23}(t)[ \beta_{2}( t) +\beta_{3}(t) ] ,  \label{13}
\end{eqnarray}%
\be
\f{\delta_{12}(t)}{d_{12}(t)} = \f{\delta_{13}(t)}{d_{13}(t)}=\f{\delta_{23}(t)}{d_{23}(t)} = F(t),
  \label{14}
\ee
where
$
F(t) = \alpha_1(t)m_1(t)
$
under the requirement
\be
\alpha_1(t)m_1(t) = \alpha_2(t)m_2(t) =\alpha_3(t)m_3(t).  \label{15}
\ee

To determine the coefficients, we first put $\alpha_{j}(t)$ as the same formula as that
we adopted in the case of two coupled oscillators \cite{tbo}:
\begin{equation}
\alpha_{j}(t)  =\alpha_{0,j}\rho_{j}^{2}(t), \label{16}%
\end{equation}
where $\rho_{j}$ are solutions of the following auxiliary equation
\begin{equation}
\ddot{\rho}_{j}+\frac{\dot{m}_{j}}{m_{j}}\dot{\rho}_{j}+\tilde{\omega}_{j}^2(t)
\rho_{j}=\frac{\Omega_{j}^2}{4m_{j}^{2}\rho_{j}^{3}},  \label{17}
\end{equation}
which are real, while $\Omega_{j}$ are real constants.
Then, $\beta_{j}(t)$ and $\gamma_{j}(t)$ are also determined like in the case of two coupled
oscillators \cite{tbo}, such that
\begin{equation}
\beta_{j}(t)=\alpha_{0,j}m_{j}(t)[b_j(t)\rho_j^2(t)- \rho_{j}(t)\dot{\rho}_{j}(t)],  \label{18}
\end{equation}%
\begin{equation}
\gamma_{j}(t)  =\alpha_{0,j} \Bigg[\f{\Omega_j^2}{4\rho_j^2(t)}
+ m_j^2(t) \Big(b_j^2(t)\rho_j^2(t)-2b_j(t)\rho_j(t)\dot{\rho}_j(t)+\dot{\rho}_j^2(t)\Big)\Bigg].
  \label{19}
\end{equation}
While, according to Eq. (\ref{14}), we can put $\delta_{jk}(t)$ as
\ba
\delta_{12}(t)&=&F(t)d_{12}(t),  \label{20} \\
\delta_{13}(t)&=&F(t)d_{13}(t),  \label{21} \\
\delta_{23}(t)&=&F(t)d_{23}(t),  \label{22}
\ea
Eqs. (\ref{11})-(\ref{13}) give the requirements that $d_{jk}(t)$ should follow.
Rigorous evaluations show
that such requirements are the relations of the form
\ba
\dot{d}_{12}(t) &=& -G_{12}(t) d_{12}(t),  \label{23}  \\
\dot{d}_{13}(t) &=& -G_{13}(t) d_{13}(t),  \label{24}   \\
\dot{d}_{23}(t) &=& -G_{23}(t) d_{23}(t),  \label{25}
\ea
where
\ba
G_{12}(t) &=& \f{\dot{m}_3(t)}{m_3(t)}+\f{\dot{\rho}_1(t)}{\rho_1(t)}+\f{\dot{\rho}_2(t)}{\rho_2(t)}+\f{2\dot{\rho}_3(t)}{\rho_3(t)},
  \label{26} \\
G_{13}(t) &=& \f{\dot{m}_2(t)}{m_2(t)}+\f{\dot{\rho}_1(t)}{\rho_1(t)}+\f{2\dot{\rho}_2(t)}{\rho_2(t)}+\f{\dot{\rho}_3(t)}{\rho_3(t)},
  \label{27} \\
G_{23}(t) &=& \f{\dot{m}_1(t)}{m_1(t)}+\f{2\dot{\rho}_1(t)}{\rho_1(t)}+\f{\dot{\rho}_2(t)}{\rho_2(t)}+\f{\dot{\rho}_3(t)}{\rho_3(t)}.
  \label{28}
\ea
The methodology of deriving $G_{jk}(t)$ is represented in Appendix A.
Thus, Eq. (\ref{6}) with Eqs. (\ref{16}), (\ref{18}), (\ref{19}), (\ref{20}), (\ref{21}),
and (\ref{22}) is the invariant operator.
This operator is valid under the two groups of conditions,
where the first group is given by Eq. (\ref{15}) and the second group
by Eqs. (\ref{23})-(\ref{25}).
Complete quantum description of coupled oscillatory systems may be possible
through the use of this dynamical invariant.
\\
\\
{\bf 3. Unitary Relations  \vspace{0.2cm}} \\
Because the formula of the invariant derived in the previous section is somewhat complicated,
its direct use in unfolding the associated quantum theory is not favorable.
Instead, developing
quantum theory of the systems with the help of the invariant operator simplified
by unitary or canonical transformations may be better.
We will adopt the unitary transformation method \cite{ada,qut,qut2,lah-3-7,ede}
among the two for that purpose in this section.
We first transform the invariant operator using a procedure
adopted in Ref. \cite{tbo} as
\be
\hat{\mathcal I}_{A}(t) = \hat{U}_{A}^{-1} \hat{\mathcal I}(t) \hat{U}_{A},  \label{29}
\ee
where $\hat{\mathcal I}_{A}(t)$ is a transformed invariant operator and
$\hat{U}_A$ is a unitary operator transforming the invariant, which reads
\begin{equation}
\hat{U}_A=\hat{U}_{A1}\hat{U}_{A2},  \label{30}
\end{equation}%
whereas
\ba
\hat{U}_{A1}&=&\prod_{j=1}^{3}\exp \left( \frac{i}{2\hbar }(\hat{p}_{j}\hat{x}_{j}
+\hat{x}_{j}\hat{p}_{j})\ln \sqrt{\f{1}{M\alpha_{j}(t)}}\right) ,
\label{31} \\
\hat{U}_{A2}&=&\exp \bigg( -\frac{i}{2\hbar }\sum_{j=1}^{3} M\beta_j( t) \hat{x}_{j}^{2}\bigg) .
\label{32}
\ea
Then, using a little bit of algebra, the invariant operator is transformed into
\ba
\hat{\mathcal I}_{A}(t) &=&\f{1}{2}\sum_{j=1}^{3}\bigg(
\f{\hat{p}_{j}^{2}}{M} + M \omega_{0,j}^2 \hat{x}_{j}^{2}\bigg) +M\delta_{12}(t)
\sqrt{\alpha_1(t)\alpha_2(t)} \hat{x}_{1}\hat{x}_{2}
\nonumber \\
& &+M\delta_{13}(t)\sqrt{\alpha_1(t)\alpha_3(t)} \hat{x}_{1}\hat{x}_{3}
+M\delta_{23}(t)\sqrt{\alpha_2(t)\alpha_3(t)} \hat{x}_{2}\hat{x}_{3},\label{33}%
\ea
where
\be
\omega_{0,j}^2 = \alpha_j(t)\gamma_{j}(t)-\beta_j^2(t)=\f{\alpha_{0,j}^2\Omega_j^2}{4}.
  \label{34}
\ee
We can also represent $\hat{\mathcal I}_{A}(t)$ in a matrix form, namely 
\ba
\hat{\mathcal I}_{A}(t) &=&\f{1}{2M}
{\bf p}^T {\bf p} + \f{1}{2}M {\bf x}^T \Gamma {\bf x} , \label{35}%
\ea
where ${\bf x}^T = (\hat{x}_1,\hat{x}_2,\hat{x}_3)$, ${\bf p}^T = (\hat{p}_1,\hat{p}_2,\hat{p}_3)$, and
\begin{equation}
\Gamma=\left(
\begin{array}{ccc}
\omega_{0,1}^2 & \Delta_{12} & \Delta_{13} \\
\Delta_{12} & \omega_{0,2}^2 & \Delta_{23} \\
\Delta_{13} & \Delta_{23} & \omega_{0,3}^2%
\end{array}%
\right),   \label{36}
\end{equation}%
while
$\Delta_{jk}=\delta_{jk}(t)\sqrt{\alpha_j(t)\alpha_k(t)}$.
From the straightforward 
evaluations of the time derivatives of $\Delta_{jk}$ using Eqs. (\ref{16}) and (\ref{20})-(\ref{22})
with Eqs. (\ref{23})-(\ref{25}), we have $d \Delta_{jk}/dt =0$.
This means that $\Delta_{jk}$ are constants.
Because $\omega_{0,j}^2$ are also constants as can be seen from Eq. (\ref{34}), all elements of
$\Gamma$ are constants.

If we denote the eigenvalues of $\Gamma$ by
$\varpi_{0,1}^2$, $\varpi_{0,2}^2$, and $\varpi_{0,3}^2$,
they are known in the literature (see Appendix B).
However, we are rather interested in the diagonalization of $\Gamma$ and
the resultant formulae of eigenvalues attained by a rotational
unitary transformation with certain angles instead of them.
This procedure is necessary for the whole
description of the unitary transformation that we have initially planed.
Attaining complete knowledge for quantum characteristics of the original systems may be possible only
through the full process of that transformation.

The transformed invariant operator, $\hat{\mathcal I}_{A}(t)$,
is simpler than the original operator, $\hat{\mathcal I}(t)$, since
the terms in the parenthesis in Eq. (\ref{33}) are identical to the Hamiltonian of SHOs.
However, $\hat{\mathcal I}_{A}(t)$
involves three coupling terms that must be removed through a further transformation.
In order to eliminate them, we consider the following transformation as the next step:
\be
\hat{\mathcal I}_{B}(t) = \hat{U}_{B}^{-1} \hat{\mathcal I}_{A}(t) \hat{U}_{B},  \label{45}
\ee
where the unitary operator $\hat{U}_B$ is of the form
\begin{equation}
\hat{U}_B=\hat{U}_{B1}\hat{U}_{B2}\hat{U}_{B3},  \label{46}
\end{equation}%
whereas
\ba
\hat{U}_{B1}&=&\exp\bigg( -\frac{i\phi}{\hbar}\left( \hat{p}_{3}\hat{x}_{2}%
-\hat{p}_{2}\hat{x}_{3}\right)  \bigg),  \label{47} \\
\hat{U}_{B2}&=&\exp\bigg( -\frac{i\theta}{\hbar}\left( \hat{p}_{1}\hat{x}_{3}%
-\hat{p}_{3}\hat{x}_{1}\right)  \bigg),  \label{48}  \\
\hat{U}_{B3}&=&\exp\bigg( -\frac{i\varphi}{\hbar}\left( \hat{p}_{2}\hat{x}_{1}%
-\hat{p}_{1}\hat{x}_{2}\right)  \bigg). \label{49}%
\ea
In fact this transformation corresponds to a rotation of the matrix formula of
$\hat{\mathcal I}_{A}(t)$ expressed in Eq. (\ref{35}):
\be
\hat{\mathcal I}_{B}(t) = \f{1}{2M}
{\bf p}^T {\bf p} + \f{1}{2}M {\bf x}^T \mathbb{R}^{T} \Gamma \mathbb{R} {\bf x},  \label{50}
\ee
where $\mathbb{R}$ is the rotation matrix that is given by (see Appendix C)
\begin{equation}
\mathbb{R} =\left(
\begin{array}{ccc}
\cos \theta \cos \varphi & -\cos \theta \sin \varphi & \sin \theta \\
\cos \phi\sin \varphi +\sin\phi\sin \theta \cos \varphi~~ & \cos \phi
\cos \varphi-\sin\phi\sin\theta\sin\varphi~~ & - \sin \phi \cos \theta
\\
\sin \phi \sin \varphi -\cos \phi \sin \theta \cos \varphi~~ & \sin \phi
\cos \varphi+\cos\phi\sin\theta\sin\varphi~~ & \cos \phi \cos \theta
\end{array}%
\right) .  \label{51}
\end{equation}%

In what follows, the transformation in Eq. (\ref{45}) results in
\be
\hat{\mathcal I}_{B}(t) =\f{1}{2}\sum_{j=1}^{3}\bigg(
\f{\hat{p}_{j}^{2}}{M} + M \bar{\omega}_{0,j}^2 \hat{x}_{j}^{2}\bigg) +M\bar{\delta}_{12}
\hat{x}_{1}\hat{x}_{2}
+M\bar{\delta}_{13} \hat{x}_{1}\hat{x}_{3}
+M\bar{\delta}_{23} \hat{x}_{2}\hat{x}_{3},\label{52}%
\ee
where
\ba
\bar{\omega}_{0,1}^2 &=& \omega_{0,1}^2\cos^2\theta\cos^2\varphi+\omega_{0,2}^2
(\sin\phi \sin\theta\cos\varphi+\cos\phi\sin\varphi)^2 \nonumber \\
& &+\omega_{0,3}^2
(\cos\phi \sin\theta\cos\varphi-\sin\phi\sin\varphi)^2
+2\{\Delta_{12}\cos\theta\cos\varphi
\nonumber \\
& &\times(\sin\phi\sin\theta\cos\varphi+\cos\phi\sin\varphi)
+\Delta_{13}\cos\theta\cos\varphi
\nonumber \\
& &\times(\sin\phi\sin\varphi-\cos\phi\sin\theta\cos\varphi)
+\Delta_{23}
[\sin\theta \cos\varphi\sin\varphi \nonumber \\
& &\times(\sin^2\phi-\cos^2\phi)
+\cos\phi\sin\phi(\sin^2\varphi-\sin^2\theta\cos^2\varphi)]
\},  \label{53} \\
\bar{\omega}_{0,2}^2 &=& \omega_{0,1}^2\cos^2\theta\sin^2\varphi+\omega_{0,2}^2
(\cos\phi\cos\varphi-\sin\phi \sin\theta\sin\varphi)^2 \nonumber \\
& &+\omega_{0,3}^2
(\sin\phi \cos\varphi+\cos\phi\sin\theta\sin\varphi)^2
+2\{\Delta_{12}\cos\theta\sin\varphi
\nonumber \\
& &\times(\sin\phi\sin\theta\sin\varphi-\cos\phi\cos\varphi)
-\Delta_{13}\cos\theta\sin\varphi
\nonumber \\
& &\times(\cos\phi\sin\theta\sin\varphi+\sin\phi\cos\varphi)
+\Delta_{23}
[\sin\theta \cos\varphi\sin\varphi \nonumber \\
& &\times(\cos^2\phi -\sin^2\phi) +\cos\phi\sin\phi(\cos^2\varphi-\sin^2\theta\sin^2\varphi)]
\},  \label{54} \\
\bar{\omega}_{0,3}^2 &=& \omega_{0,1}^2\sin^2\theta+\omega_{0,2}^2
\sin^2\phi \cos^2\theta+\omega_{0,3}^2
\cos^2\phi \cos^2\theta
 \nonumber \\
& &-2\{\Delta_{12}\sin\phi\sin\theta\cos\theta
-\Delta_{13} \cos\phi\cos\theta\sin\theta
\nonumber \\
& &+\Delta_{23}
\cos\phi\sin\phi\cos^2\theta \}.  \label{55}
\ea
By the way, we have represented the formulae of $\bar{\delta}_{jk}$ appeared in Eq. (\ref{52}), separately,
in Appendix D, since they are much more complicated and not so essential.
Now the matrix formula of $\hat{\mathcal I}_{B}( t)$ becomes
\ba
\hat{\mathcal I}_{B}(t) &=&\f{1}{2M}
{\bf p}^T {\bf p} + \f{1}{2}M {\bf x}^T \bar{\Gamma} {\bf x} , \label{56}%
\ea
where
\begin{equation}
\bar{\Gamma}=\left(
\begin{array}{ccc}
\bar{\omega}_{0,1}^2 & \bar{\delta}_{12} & \bar{\delta}_{13} \\
\bar{\delta}_{12} & \bar{\omega}_{0,2}^2 & \bar{\delta}_{23} \\
\bar{\delta}_{13} & \bar{\delta}_{23} & \bar{\omega}_{0,3}^2%
\end{array}%
\right).   \label{57}
\end{equation}%

For the purpose of diagonalization of Eq. (\ref{57}), we take angles as
\begin{eqnarray}
\phi &=& {\rm atan} (u_\phi, v_\phi),     \label{58} \\
\theta &=& {\rm atan} (u_\theta, v_\theta), \label{59} \\
\varphi &=& \pm {\rm atan}  (u_\varphi, v_\varphi),   \label{60}
\end{eqnarray}%
where $\vartheta \equiv {\rm atan} (z_1, z_2)$ is the two-variable arctangent function of $\tan \vartheta = z_2/z_1$, and
\ba
u_\phi &=&2(\varpi_{0,1}^2-\varpi_{0,3}^2)(\varpi_{0,2}^2-\varpi_{0,3}^2)
(\varpi_{0,3}^2-\omega_{0,1}^2)\Delta_{13} \sin\theta
-(\varpi_{0,1}^2-\varpi_{0,2}^2)  \nonumber \\
& &\times[(\varpi_{0,1}^2+\varpi_{0,2}^2)(\varpi_{0,3}^2-\omega_{0,1}^2)
-\varpi_{0,3}^4+\omega_{0,1}^4+ \Delta_{12}^2+\Delta_{13}^2]\Delta_{12}
\sin(2\varphi),  \label{61} \\
v_\phi &=&2(\varpi_{0,1}^2-\varpi_{0,3}^2)(\varpi_{0,2}^2-\varpi_{0,3}^2)
(\omega_{0,1}^2-\varpi_{0,3}^2)\Delta_{12} \sin\theta
-(\varpi_{0,1}^2-\varpi_{0,2}^2)  \nonumber \\
& &\times[(\varpi_{0,1}^2+\varpi_{0,2}^2)(\varpi_{0,3}^2-\omega_{0,1}^2)
-\varpi_{0,3}^4+\omega_{0,1}^4+ \Delta_{12}^2+\Delta_{13}^2]\Delta_{13}
\sin(2\varphi),  \label{62} \\
u_\theta &=&[(\varpi_{0,3}^2-\omega_{0,1}^2)(\varpi_{0,3}^2-\varpi_{0,1}^2-\varpi_{0,2}^2
+\omega_{0,1}^2)- \Delta_{12}^2-\Delta_{13}^2]^{1/2},  \label{63} \\
v_\theta &=&[\omega_{0,1}^4-(\varpi_{0,1}^2+\varpi_{0,2}^2)\omega_{0,1}^2
+\varpi_{0,1}^2 \varpi_{0,2}^2+ \Delta_{12}^2+\Delta_{13}^2]^{1/2},  \label{64} \\
u_\varphi &=&\{(\varpi_{0,2}^2-\varpi_{0,3}^2)
[\omega_{0,1}^4+\varpi_{0,2}^2(\varpi_{0,3}^2-\omega_{0,1}^2)-\varpi_{0,3}^2\omega_{0,1}^2
+\Delta_{12}^2+\Delta_{13}^2]\}^{1/2},  \label{63} \\
v_\varphi &=&-\{(\varpi_{0,3}^2-\varpi_{0,1}^2)
[\omega_{0,1}^4+\varpi_{0,1}^2(\varpi_{0,3}^2-\omega_{0,1}^2)-\varpi_{0,3}^2\omega_{0,1}^2
+\Delta_{12}^2+\Delta_{13}^2]\}^{1/2},  \label{65}
\ea
while $\varpi_{0,j}^2$ are given in Appendix B.
The function ${\rm atan} (z_1, z_2)$ is defined during one cycle:
for instance, it is defined in the range $-\pi < \vartheta \leq \pi$
in Mathematica program \cite{swo}.
There also exist
other diagonalization sets of angles instead of Eqs. (\ref{58})-(\ref{60}), and
we have represented them in Appendix E.

There are two categories of the matrix $\Gamma$ in this context, where the plus sign in Eq. (\ref{60}) is applied
to the first category (class 1) whereas the minus sign to the second category (class 2).
To see the details
of the two classes of $\Gamma$, let us look the transformation $\mathbb{R}^{T} \Gamma \mathbb{R}$ in Eq. (\ref{50}),
which can be fulfilled in relation with $\phi$, $\theta$, and $\varphi$ in turn using Eq. (\ref{C1})
in Appendix C.
We consider the transformation up to $\theta$ in this process:
\be
\bar{\Gamma}_{\theta} = \mathbb{R}_{x_2}^{T}(\theta) \mathbb{R}_{x_1}^{T}(\phi) \Gamma \mathbb{R}_{x_1}(\phi)\mathbb{R}_{x_2}(\theta),
\label{theta}
\ee
where $\phi$ is being expressed in terms of $\varphi$ using Eq. (\ref{60}) with the plus sign.
Then $\Gamma$ is the first category if and only if this procedure
yields $\bar{\delta}_{13}'=\bar{\delta}_{23}'=0$,
where $\bar{\delta}_{13}'$ ($\bar{\delta}_{23}'$) is an element of $\bar{\Gamma}_{\theta}$,
which corresponds to the first row and third column (the second row and third column);
$\Gamma$ is the second category otherwise.
This manifestation is the definition of the two classes of $\Gamma$ or the rule for
distinguishing them from each other.
In fact, for the case of class 2, the transformation, Eq. (\ref{theta}), with the choice of $\phi$
represented in terms of $\varphi$ in Eq. (\ref{60}) with the minus sign
gives $\bar{\delta}_{13}'=\bar{\delta}_{23}'=0$.

We see that all $\bar{\delta}_{jk}$
in Appendix D reduce to zero by choosing angles as Eqs. (\ref{58})-(\ref{60}),
leading to attaining the diagonalization of $\bar{\Gamma}$.
Meanwhile, the momentum parts in Eq. (\ref{52}) do not altered by this transformation.
Thus the finally transformed invariant is just written as
\be
\hat{\mathcal I}_{B} =\f{1}{2}\sum_{j=1}^{3}\bigg(
\f{\hat{p}_{j}^{2}}{M} + M \bar{\omega}_{0,j}^2 \hat{x}_{j}^{2}\bigg) . \label{68}%
\ee
Although the formulae of $\bar{\omega}_{0,j}^2$ in this equation
are somewhat complicated as can be seen from Eqs. (\ref{53})-(\ref{55}),
they are constants over time because they are represented in terms of  $\omega_{0,j}^2$ and $\Delta_{jk}$
only, which are already proved to be constants.
$\bar{\omega}_{0,j}^2$ are mathematically equivalent to $\varpi_{0,j}^2$
given in Appendix B respectively, since the considered rotational angles,
Eqs. (\ref{58})-(\ref{60}), are eigenangles.
For actual cases, it is better to treat the transformed systems by replacing $\bar{\omega}_{0,j}^2$ with
$\varpi_{0,j}^2$ in Eq. (\ref{68}) because $\varpi_{0,j}^2$ are much simpler in a relative sense.
The matrix, Eq. (\ref{36}), is positive-definite when and only when all of the leading principal minors are positive
according to the Sylvester's criterion \cite{pdm}.
For more detailed descriptions of the condition for the existence of
such positive-definite eigenvalues, refer for example to Ref. \cite{cpl}.
\\
\\
{\bf 4. Quantum Solutions \vspace{0.2cm}} \\
We will show in this section that the previous formulation of the invariant and the related unitary relations can be
utilized to derive quantum solutions of the systems.
We introduce annihilation operators associated with SHOs for that purpose, such that
\be
\hat{a}_{0,j} = \sqrt{\f{M \bar{\omega}_{0,j}}{2\hbar}} \hat{x}_j + \f{i}{\sqrt{2M\bar{\omega}_{0,j}\hbar}}\hat{p}_j,
  \label{69}
\ee
and the corresponding creation operators as the Hermitian adjoint of Eq. (\ref{69}), $\hat{a}_{0,j}^\dagger$.
Then, it is possible to represent Eq. (\ref{68}) in the form
\be
\hat{\mathcal I}_{B} =\sum_{j=1}^{3}\hbar \bar{\omega}_{0,j} \bigg(\hat{a}_{0,j}^\dagger \hat{a}_{0,j} + \f{1}{2}\bigg).
  \label{70}
\ee
The annihilation operators, $\hat{a}_j$, in the original systems are related to $\hat{a}_{0,j}$ by
\be
\hat{a}_j = \hat{U}_{A} \hat{U}_{B}\hat{a}_{0,j}\hat{U}_{B}^{-1}\hat{U}_{A}^{-1}.  \label{71}
\ee
The complete formulae of $\hat{a}_j$ are shown in Appendix F.
Now we can express
the invariant operator in the original systems in terms of $\hat{a}_j$ and
their Hermitian adjoints $\hat{a}_j^\dagger$ (creation operators):
\be
\hat{\mathcal I}(t) =\sum_{j=1}^{3}\hbar \bar{\omega}_{0,j} \bigg(\hat{a}_j^\dagger \hat{a}_j + \f{1}{2}\bigg).
  \label{72}
\ee

Let us write the eigenvalue equations for the lastly transformed invariant operator as
\be
\hat{\mathcal I}_{B} u_{0,n_{1},n_{2},n_{3}}(x_1,x_2,x_3) = \lambda_{n_{1},n_{2},n_{3}} u_{0,n_{1},n_{2},n_{3}}(x_1,x_2,x_3),  \label{73}
\ee
where $\lambda_{n_{1},n_{2},n_{3}}$ are eigenvalues and $u_{0,n_{1},n_{2},n_{3}}(x_1,x_2,x_3)$ are eigenfunctions.
$\lambda_{n_{1},n_{2},n_{3}}$ are constants because $\hat{\mathcal I}_{B}$ is independent of time.
By solving Eq. (\ref{73}), we have the familiar eigenfunctions and eigenvalues which are
\begin{equation}
u_{0,n_{1},n_{2},n_{3}}(x_1,x_2,x_3) = \prod_{j=1}^{3}
\sqrt[4]{\frac{M\bar{\omega}_{0,j}}{\pi\hbar }}
\f{1}{\sqrt{2^{n_j}n_j!}}
H_{n_j}\left(  \sqrt{\frac{M\bar{\omega}_{0,j}}{\hbar}%
}x_j\right)  \exp\left[  -\frac{M\bar{\omega}_{0,j}}{2\hbar}x_j^{2}\right]  ,  \label{74}
\end{equation}
\be
\lambda_{n_{1},n_{2},n_{3}} =\sum_{j=1}^{3}\hbar \bar{\omega}_{0,j} \bigg(n_j + \f{1}{2}\bigg).
  \label{75}
\ee
The eigenvalue equations in the original systems can also be written as
\be
\hat{\mathcal I}(t) u_{n_{1},n_{2},n_{3}}(x_1,x_2,x_3,t) = \lambda_{n_{1},n_{2},n_{3}} u_{n_{1},n_{2},n_{3}}(x_1,x_2,x_3,t),  \label{76}
\ee
where $u_{n_{1},n_{2},n_{3}}(x_1,x_2,x_3,t)$ are eigenfunctions, while the eigenvalues are the same as Eq. (\ref{75}).

The mathematical relation between the eigenfunctions in the original systems and those in the transformed systems
is given by
\be
u_{n_{1},n_{2},n_{3}}(x_1,x_2,x_3,t) = \hat{U} u_{0, n_{1},n_{2},n_{3}}(x_1,x_2,x_3),  \label{77}
\ee
where
\be
\hat{U} = \hat{U}_{A} \hat{U}_{B}. \label{78}
\ee
Note that this relation is inverse of the previous transformation.
By evaluating Eq. (\ref{77}), we easily have
\begin{eqnarray}
u_{n_{1},n_{2},n_{3}}(x_1,x_2,x_3,t) &=& \prod_{j=1}^{3}
\sqrt[4]{\frac{\bar{\omega}_{0,j}}{\pi\hbar\alpha_j(t) }}
\f{1}{\sqrt{2^{n_j}n_j!}}
H_{n_j}\left(  \sqrt{\frac{\bar{\omega}_{0,j}}{\hbar}%
}X_j\right) \nonumber \\
& &\times\exp\left[  -\frac{1}{2\hbar}\bigg(\bar{\omega}_{0,j}X_j^{2}+i \f{\beta_j(t)}{\alpha_j(t)}x_j^{2} \bigg)\right]  ,  \label{79}
\end{eqnarray}
where
\begin{equation}
\left(
\begin{array}
[c]{c}%
X_{1}\\
X_{2} \\
X_3
\end{array}
\right)  = \mathbb{R}^T \left(
\begin{array}
[c]{c}%
\alpha_1^{-1/2}(t) x_{1}\\
\alpha_2^{-1/2}(t) x_{2} \\
\alpha_3^{-1/2}(t) x_3
\end{array}
\right)  . \label{80}%
\end{equation}
Hence, the eigenfunctions of the complicated original invariant operator are derived by
inversely transforming
the simple eigenfunctions associated with the transformed invariant operator.
These eigenfunctions are basic in the study of quantum features of the systems.
\\
\\
{\bf 5. Conclusion
\vspace{0.2cm}}
\\
A general dynamical invariant operator of time-dependent three coupled oscillators was formulated
based on its mathematical definition.
The parameters of the oscillatory systems that we have considered vary in the most general way in
time so long as the restrictions raised in the invariant formulation allow.
The invariant operator was diagonalized by its unitary transformations.
From such a procedure, the unitary relation
between the original invariant operator and the one for SHOs was elucidated.

The transformation of the invariant was performed in two steps.
The invariant operator was simplified by the first transformation as can be seen from Eq. (\ref{33}),
but it still involves three cross terms.
Through the second transformation, the cross terms have been removed and, as a consequence,
the invariant operator has been represented in terms of constant parameters only.

Using the fact that the eigenfunctions of the transformed (or diagonalized) invariant operator
are well known, we obtained the eigenfunctions associated with
the original systems via the inverse transformation of such known ones.
Our analysis in this work is exact provided that
the two groups of conditions in parameteric variations given in the text hold.
In contrast to this, approximations have been usually employed in the
previous analyses of time-dependent coupled oscillators,
under the assumption of the adiabatic evolution of the systems \cite{lah-3-7,ede} or
sudden quenches of parameters \cite{ada,sgh,DP1,DP2}.
Some authors otherwise considered particular cases where the additional term
does not appear in the transformation of the Hamiltonian or at least it can be neglected \cite{VERSION1,mg}.

It may be possible to use our analysis of the dynamical invariant in characterizing
quantum properties of coupled oscillatory systems, such as
nano-optomechanical systems \cite{aa1,aa2,ede},
arrays of electromechanical devices \cite{aed}, and biological/neural oscillator networks \cite{bnn}.
According to the recent trend that the size of optomechanical and electromechanical devices
becomes smaller and smaller towards nanoscale,
the quantum features in such systems become prominent.
Our dynamical invariant developed in this work is crucial as a tool of quantum analyses of them,
because a large part of such devices are
described by using a model of coupled oscillators.

\appendix
\section{\bf About the formula of $G_{jk}(t)$}
Equations (\ref{23})-(\ref{25}) which involve
$G_{jk}(t)$ are evaluated using Eqs. (\ref{11})-(\ref{13}), respectively.
We 
see from Eq. (\ref{15}) that $F(t)$
can be represented in three other forms. Among them, $F(t)=\alpha_3(t)m_3(t)$ is
used when we derive Eq. (\ref{23}) from Eq. (\ref{11}),
$F(t)=\alpha_2(t)m_2(t)$ is used when we derive Eq. (\ref{24}) from Eq. (\ref{12}),
and $F(t)=\alpha_1(t)m_1(t)$
is used when we derive Eq. (\ref{25}) from Eq. (\ref{13}).

However, other combinations of the formulae of $F(t)$ are equally allowed in the derivations of the
three 
$G_{jk}(t)$.
For instance, if we use $F(t)=\alpha_1(t)m_1(t)$ for all three
derivations of Eqs. (\ref{23})-(\ref{25}), we have.
\ba
G_{12}(t) &=& \f{\dot{m}_1(t)}{m_1(t)}+\f{3\dot{\rho}_1(t)}{\rho_1(t)}+\f{\dot{\rho}_2(t)}{\rho_2(t)},  \label{A1} \\
G_{13}(t) &=& \f{\dot{m}_1(t)}{m_1(t)}+\f{3\dot{\rho}_1(t)}{\rho_1(t)}+\f{\dot{\rho}_3(t)}{\rho_3(t)},  \label{A2} 
\ea
whereas $G_{23}(t)$ is already given in Eq. (\ref{25}) in the text (i.e., it is not altered).
The formulae of $G_{jk}(t)$ obtained with the use of other combinations of the formulae of $F(t)$ can now
be easily conjectured  through the expressions 
given in Eqs. (\ref{A1}) and (\ref{A2}), and other expressions along this line
in the text.

\section{\bf The eigenvalues of $\Gamma$}
The eigenvalues $\varpi_{0,j}^{2}$ of $\Gamma$ appeared in Eq. (\ref{36}) have been
reported in previous literature \cite{dpt,dpt2}.
They are given by
\begin{eqnarray}
\varpi_{0,1}^{2} &=& \f{1}{3}\bigg[\omega_0^2 + \frac{J }{\sqrt{2}}\cos \Theta \bigg] ,
  \label{37} \\
\varpi_{0,2}^{2} &=& \f{1}{3}\bigg[\omega_0^2 + \frac{J }{\sqrt{2}}\cos \left( \Theta -\frac{2\pi
}{3}\right) \bigg] ,  \label{38} \\
\varpi_{0,3}^{2} &=& \f{1}{3}\bigg[\omega_0^2 + \frac{J }{\sqrt{2}}\cos \left( \Theta +\frac{2\pi
}{3}\right) \bigg] ,  \label{39}
\end{eqnarray}%
where $\omega_0 = (\omega_{0,1}^{2}+\omega_{0,2}^{2}+\omega_{0,3}^{2})^{1/2}$ and%
\begin{equation}
J =2\Big[( \omega_{0,1}^{2}-\omega_{0,2}^{2}) ^{2}
+( \omega_{0,1}^{2}-\omega_{0,3}^{2}) ^{2}+( \omega _{0,2}^{2}
-\omega_{0,3}^{2}) ^{2}+6\Delta^2 \Big]^{1/2},  \label{40}
\end{equation}
\begin{equation}
\Theta =\frac{1}{3}\arccos \left( \frac{A}{2 B^{3/2}}\right) ,  \label{41}
\end{equation}%
with%
\begin{eqnarray}
\Delta &=& (\Delta_{12}^{2}+\Delta_{13}^{2}+\Delta_{23}^{2})^{1/2},  \label{42} \\
A &=&-3( \omega_{0,1}^{2}+\omega_{0,2}^{2})
( \omega_{0,1}^{2}+\omega_{0,3}^{2}) ( \omega_{0,2}^{2}+\omega_{0,3}^{2})
\nonumber \\
&&-27\left( \omega_{0,1}^{2}\Delta_{23}^{2}+\omega_{0,2}^{2}%
\Delta_{13}^{2}+\omega_{0,3}^{2}\Delta_{12}^{2}\right)+9\omega_0^2 \Delta^2
\nonumber \\
&&+2( \omega_{0,1}^{6}+\omega_{0,2}^{6}+\omega_{0,3}^{6}) +18\left(
\omega_{0,1}^{2}\omega_{0,2}^{2}\omega_{0,3}^{2}+3\Delta_{12}\Delta_{13}\Delta_{23}%
\right) ,  \label{43} \\
B &=&\frac{1}{2}\Big[ ( \omega_{0,1}^{2}-\omega_{0,2}^{2})
^{2}+( \omega_{0,1}^{2}-\omega_{0,3}^{2}) ^{2}
+( \omega_{0,2}^{2}-\omega_{0,3}^{2}) ^{2}\Big] +3\Delta^2 .  \label{44}
\end{eqnarray}%
Note that the magnitudes of $\varpi_{0,i}^{2}$ are 
in decreasing order:
$\varpi_{0,1}^{2} \geq \varpi_{0,2}^{2} \geq \varpi_{0,3}^{2}$.

\section{\bf About the formula of $\mathbb{R}$}
The matrix $\mathbb{R}$ given in Eq. (\ref{51}) can be rewritten by rotation matrices of each angle as \cite{VERSION1}
\begin{eqnarray}
\mathbb{R}&=& \mathbb{R}_{x_1} (\phi) \mathbb{R}_{x_2} (\theta) \mathbb{R}_{x_3} (\varphi) \nonumber \\
& &=
\left(
\begin{array}{ccc}
1 & 0 & 0 \\
0 & \cos \phi & -\sin \phi \\
0 & \sin \phi & \cos \phi%
\end{array}%
\right)
\left(
\begin{array}{ccc}
\cos \theta & 0 & \sin \theta \\
0 & 1 & 0 \\
-\sin \theta & 0 & \cos \theta%
\end{array}%
\right)
\left(
\begin{array}{ccc}
\cos \varphi & -\sin \varphi & 0 \\
\sin \varphi & \cos \varphi & 0 \\
0 & 0 & 1%
\end{array}%
\right) . \label{C1}
\end{eqnarray}%

\section{\bf The representation of $\bar{\delta}_{jk}$}
The mathematical representations of $\bar{\delta}_{jk}$ appeared in Eq. (\ref{52}) are
\ba
\bar{\delta}_{12} &=& -\omega_{0,1}^2\cos^2\theta\cos\varphi\sin\varphi
+\omega_{0,2}^2 [\cos\phi\sin\phi\sin\theta\cos(2\varphi)
\nonumber \\
& &+\cos\varphi\sin\varphi
(\cos^2\phi
-\sin^2\phi\sin^2\theta)]
-\omega_{0,3}^2 [\cos\phi\sin\phi\sin\theta\cos(2\varphi)
\nonumber \\
& &+\cos\varphi\sin\varphi(\cos^2\phi\sin^2\theta
-\sin^2\phi)]
+\Delta_{12}\cos\theta [\cos\phi\cos(2\varphi)
\nonumber \\
& &-2\sin\phi\sin\theta\cos\varphi\sin\varphi]
+\Delta_{13}\cos\theta [\sin\phi(\cos^2\varphi
-\sin^2\varphi) \nonumber \\
& &+\cos\phi\sin\theta\sin(2\varphi)]
+\Delta_{23}
\{\sin\theta\cos(2\varphi)(\sin^2\phi
-\cos^2\phi)
\nonumber \\
& &+(1/4)[3-\cos(2\theta)]\sin(2\phi)\sin(2\varphi)\} ,  \label{D1}
\\
\bar{\delta}_{13} &=& \omega_{0,1}^2\cos\theta\sin\theta\cos\varphi
-\omega_{0,2}^2\sin\phi\cos\theta (\sin\phi\sin\theta\cos\varphi
\nonumber \\
& &+\cos\phi\sin\varphi)
+\omega_{0,3}^2 \cos\phi\cos\theta(\sin\phi\sin\varphi
-\cos\phi\sin\theta\cos\varphi)
\nonumber \\
& &-\Delta_{12}[\sin\phi\cos^2\theta \cos\varphi
-\sin\theta(\sin\phi\sin\theta\cos\varphi+\cos\phi\sin\varphi)]
\nonumber \\
& &+\Delta_{13}[\cos\phi\cos^2\theta \cos\varphi
+\sin\theta(\sin\phi\sin\varphi-\cos\phi\sin\theta\cos\varphi)]
\nonumber \\
& &+\Delta_{23}
\cos\theta[\sin(2\phi)\sin\theta\cos\varphi
+\cos(2\phi)\sin\varphi] ,  \label{D2}
\\
\bar{\delta}_{23} &=& -\omega_{0,1}^2\cos\theta\sin\theta\sin\varphi
+\omega_{0,2}^2\sin\phi\cos\theta (\sin\phi\sin\theta\sin\varphi
\nonumber \\
& &-\cos\phi\cos\varphi)
+\omega_{0,3}^2 \cos\phi\cos\theta(\sin\phi\cos\varphi
+\cos\phi\sin\theta\sin\varphi)
\nonumber \\
& &+\Delta_{12}[\cos\phi\sin\theta \cos\varphi
+\sin\phi\cos(2\theta)\sin\varphi]
\nonumber \\
& &+\Delta_{13}[\sin\phi\sin\theta \cos\varphi
-\cos\phi\cos(2\theta)\sin\varphi]
\nonumber \\
& &+\Delta_{23}
\cos\theta[\cos\varphi(\cos^2\phi
-\sin^2\phi)-\sin(2\phi)\sin\theta\sin\varphi] .  \label{D3}
\ea

\section{\bf Diagonalization angles}
The matrix $\Gamma$ can also be diagonalized by other angle sets instead of Eqs. (\ref{58})-(\ref{60}).
For instance, it is diagonalized by the unitary or matrix transformation using the following sets of angles $[\phi, \theta,\varphi]$:
\ba
& &[{\rm atan} (u_\phi, v_\phi),~ \mp {\rm atan} (u_\theta, v_\theta),~ {\rm atan} (u_\varphi, -v_\varphi)], \\
& &[{\rm atan} (u_\phi, v_\phi),~ \pm {\rm atan} (u_\theta, v_\theta),~ -{\rm atan} (u_\varphi, -v_\varphi)], \\
& &[{\rm atan} (u_\phi, v_\phi),~ {\rm atan} (u_\theta, v_\theta),~ \pm {\rm atan} (-u_\varphi, -v_\varphi)], \\
& &[{\rm atan} (u_\phi, v_\phi),~ {\rm atan} (u_\theta, -v_\theta),~ {\rm atan} (u_\varphi, \mp v_\varphi)], \\
& &[{\rm atan} (u_\phi, v_\phi),~ {\rm atan} (-u_\theta, v_\theta),~ {\rm atan} (\mp u_\varphi, -v_\varphi)],
\ea
where upper signs are for the class 1 of $\Gamma$ and lower signs for class 2.
One can diagonalize
$\Gamma$ using a set of angles among six sets (the one in Eqs. (\ref{58})-(\ref{60})
and the above five sets) or another set unknown yet, depending on one's taste.
However, there is no set of angles that can diagonalize both classes of $\Gamma$.
One should be careful that the first transformation for class 1 (class 2) of $\Gamma$ must
be carried out with respect to $\phi$ represented in terms of $\theta$ and $\varphi$ with
upper (lower) signs for all six sets.

\section{\bf Full representation of $\hat{a}_j$}
The straightforward evaluation of Eq. (\ref{71}) using Eqs. (\ref{30}) and (\ref{46}) gives
\be
\hat{a}_{j} = \sqrt{\f{\bar{\omega}_{0,j}}{2\hbar}} \hat{X}_j + \f{i}{\sqrt{2\bar{\omega}_{0,j}\hbar}}\hat{P}_j,
  \label{}
\ee
where
\begin{eqnarray}
\left(
\begin{array}
[c]{c}%
\hat{X}_{1}\\
\hat{X}_{2} \\
\hat{X}_3
\end{array}
\right)  &=& \mathbb{R}^T \left(
\begin{array}
[c]{c}%
\alpha_1^{-1/2}(t) \hat{x}_{1}\\
\alpha_2^{-1/2}(t) \hat{x}_{2} \\
\alpha_3^{-1/2}(t) \hat{x}_3
\end{array}
\right) ,
\\
\left(
\begin{array}
[c]{c}%
\hat{P}_{1}\\
\hat{P}_{2} \\
\hat{P}_3
\end{array}
\right)  &=& \mathbb{R}^T \left(
\begin{array}
[c]{c}%
\alpha_1^{1/2}(t) \{\hat{p}_{1} + [\beta_1(t)/\alpha_1(t)]\hat{x}_{1} \} \\
\alpha_2^{1/2}(t) \{\hat{p}_{2} + [\beta_2(t)/\alpha_2(t)]\hat{x}_{2} \} \\
\alpha_3^{1/2}(t) \{\hat{p}_{3} + [\beta_3(t)/\alpha_3(t)]\hat{x}_{3} \}
\end{array}
\right)
. \label{}%
\end{eqnarray}
\\
\\



\begin{references}

\bibitem{pce} E. Takou, E. Barnes, and S. E. Economou,
Precise control of entanglement in multinuclear spin registers coupled to defects.
arXiv:2203.09459v2 [quant-ph] (2022).

\bibitem{pce2} W. Dong, F. A. Calderon-Vargas, and S. E. Economou,
Precise high-fidelity electron-nuclear spin entangling gates in NV centers via hybrid
dynamical decoupling sequences.
{\it New J. Phys.} {\bf 22}, 073059 (2020).

\bibitem{pce3} C. E. Bradley, J. Randall, M. H. Abobeih, R. C. Berrevoets, M. J. Degen,
M. A. Bakker, M. Markham, D. J. Twitchen, and T. H. Taminiau,
A ten-qubit solid-state spin register with quantum memory up to one minute.
{\it Phys. Rev. X} {\bf 9}(3), 031045 (2019).

\bibitem{qde} G. Csaba and W. Porod,
Coupled oscillators for computing: a review and perspective.
{\it Appl. Phys. Rev.} {\bf 7}(1), 011302 (2020).

\bibitem{qde2} K. Komarova, H. Gattuso, R. D. Levine, and F. Remacle,
Quantum device emulates the dynamics of two coupled oscillators.
{\it J. Phys. Chem. Lett.} {\bf 11}(17), 6990--6995 (2020).

\bibitem{qde3} A. Mallick, M. K. Bashar, D. S. Truesdell, B. H. Calhoun, S. Joshi, and N. Shukla,
Using synchronized oscillators to compute the maximum independent set.
{\it Nat Commun.} {\bf 11}, 4689 (2020).

\bibitem{ab} A. G. Litvak and M. D. Tokman,
Electromagnetically induced transparency in ensembles of classical oscillators.
{\it Phys. Rev. Lett.}
{\bf 88}(9), 095003 (2002).

\bibitem{ab2} C. L. G. Alzar, M. A. G. Martinez, and P. Nussenzveig,
Classical analog of electromagnetically induced transparency.
{\it Am. J. Phys.} {\bf 70}(1), 37--41 (2002).

\bibitem{saa} Y. Muraki,
Application of a coupled harmonic oscillator model to solar activity and El Ni\~{n}o phenomena.
{\it J. Astron. Space Sci.} {\bf 35}(2), 75--81 (2018).

\bibitem{lgi} S. Dutta, A. Parihar, A. Khanna, J. Gomez, W. Chakraborty, M. Jerry, B. Grisafe, A. Raychowdhury, and S. Datta,
Programmable coupled oscillators for synchronized locomotion.
{\it Nat. Commun.} {\bf 10}, 3299 (2019).

\bibitem{aot} P. S. Stein,
Application of the mathematics of coupled oscillator systems to the analysis of the neural control of locomotion.
{\it Fed. Proc.} {\bf 36}(7), 2056--2059 (1977).

\bibitem{csl} G. C. Dente, C. E. Moeller, and P. S. Durkin,
Coupled oscillators at a distance: applications to coupled semiconductor lasers.
{\it IEEE J. Quantum Electron.} {\bf 26}(6), 1014--1022 (1990).

\bibitem{VERSION1} S. Hassoul, S. Menouar, H. Benseridi, and J. R. Choi,
Quantum dynamics for general time-dependent three coupled oscillators based on an exact decoupling.
{\it Physica A} {\bf 604}, 127755 (2022).

\bibitem{ada} R. Habarrih, A. Jellal, and A. Merdaci,
Dynamics and redistribution of entanglement and coherence in three time-dependent coupled harmonic oscillators.
{\it Int. J. Geom. Methods Mod. Phys.} {\bf 18}(8), 2150120 (2021).

\bibitem{Lewis1} H. R. Lewis, Jr.,
Class of exact invariants for classical and quantum time-dependent harmonic oscillators.
\textit{J. Math. Phys.} {\bf 9}(11), 1976--1986 (1968).

\bibitem{Lewis2} H. R. Lewis, Jr. and W. B. Riesenfeld,
An exact quantum theory of the time-dependent harmonic oscillator and of a
charged particle in a time-dependent electromagnetic field.
\textit{J. Math. Phys}. {\bf 10}(8), 1458--1473 (1969).

\bibitem{tbo} J. R. Choi,
Formulation of general dynamical invariants and their unitary relations for time-dependent coupled quantum oscillators.
arXiv:2210.07551v1 [quant-ph] (2022).

\bibitem{do2} T. J. Li,
A concise quantum mechanical treatment of the forced damped harmonic oscillator.
{\it Cent. Eur. J. Phys.} {\bf 6}(4), 891--894 (2008).

\bibitem{tla} R. Daneshmand and M. K. Tavassoly,
Description of atom-field interaction via quantized Caldirola-Kanai Hamiltonian.
{\it Int. J. Theor. Phys.} {\bf 56}(4), 1218--1232 (2017).

\bibitem{do1} D. Chru\'{s}ci\'{n}ski and J. Jurkowski,
Quantum damped oscillator I: Dissipation and resonances.
{\it Ann. Phys.} {\bf 321}(4), 854--874 (2006).

\bibitem{do1-1} D. Chru\'{s}ci\'{n}ski,
Quantum damped oscillator II: Bateman’s Hamiltonian vs. 2D parabolic potential barrier.
{\it Ann. Phys.} {\bf 321}(4), 840--853 (2006).

\bibitem{qut} K. H. Yeon, C. I. Um, S.-K. Hong, and T. F. George,
Quantum unitary transformation corresponding to the classical square canonical transformation and
its connected quantum systems.
{\it J. Korean Phys. Soc.} {\bf 46}(3), 591--596 (2005).

\bibitem{qut2} Z.-Z. Li, W.-H. Han, and Z.-Y. Li,
Unitary transformation of general nonoverlapping-image multimode interference couplers
with any input and output ports.
{\it Chin. Phys. B} {\bf 29}(1), 014206 (2020).

\bibitem{lah-3-7} J. R. Choi and S. Ju,
Quantum characteristics of a nanomechanical resonator coupled to a superconducting LC resonator
in quantum computing systems.
{\it Nanomaterials} {\bf 9}(1), 20 (2019).

\bibitem{ede} J. R. Choi,
Entropic analysis of optomechanical entanglement for a nanomechanical resonator coupled to
an optical cavity field.
{\it SciPost Phys. Core} {\bf 4}(3), 024 (2021).

\bibitem{swo} S. Wolfram, The Mathematica Book (Wolfram Media, Champaign, 2003), 5th ed.

\bibitem{pdm} G. T. Gilbert,
Positive definite matrices and Sylvester's criterion.
{\it Am. Math. Mon.} {\bf 98}(1), 44--46 (1991).

\bibitem{cpl} F. M. Fern\'{a}ndez,
Comment on: \textquotedblleft Entanglement in three coupled oscillators" [Phys. Lett. A 384 (2020) 126134].
{\it Phys. Lett. A} {\bf 384}, 126577 (2020).

\bibitem{DP1} D.-K. Park,
Dynamics of entanglement and uncertainty relation in coupled harmonic
oscillator system: exact results.
\textit{Quantum Inf. Process}. {\bf17}(6), 147 (2018).

\bibitem{sgh} S. Ghosh, K. S. Gupta, and S. C. L. Srivastava, Entanglement dynamics following a sudden quench:
An exact solution.
{\it Europhys. Lett.} {\bf 120}(5), 50005 (2017).

\bibitem{DP2} D.-K. Park, Dynamics of entanglement in three coupled harmonic oscillator
system with arbitrary time-dependent frequency and coupling constants.
{\it Quantum Inf. Process.} {\bf 18}(9), 282 (2019).

\bibitem{mg} D. X. Macedo and I. Guedes,
Time-dependent coupled harmonic oscillators.
{\it J. Math. Phys.} {\bf 53}(5), 052101 (2012).

\bibitem{aa1} S. Chakraborty and A. K. Sarma,
Entanglement dynamics of two coupled mechanical oscillators in modulated optomechanics.
{\it Phys. Rev. A} {\bf 97}(2), 022336 (2018).

\bibitem{aa2} M. H. Nadiki and M. K. Tavassoly,
The amplitude of the cavity pump field and dissipation effects on the entanglement dynamics
and statistical properties of an optomechanical system.
{\it Opt. Commun.} {\bf 452}(5), 31--39 (2019).

\bibitem{aed} I. Mahboob, M. Mounaix, K. Nishiguchi, A. Fujiwara, and H. Yamaguchi,
A multimode electromechanical parametric resonator array.
{\it Sci. Rep.} {\bf 4}, 4448 (2014).

\bibitem{bnn} C. Bick, M. Goodfellow, C. R. Laing, and E. A. Martens,
Understanding the dynamics of biological and neural oscillator networks through exact mean-field reductions: a review.
{\it J. Math. Neurosci.} {\bf 10}, 9 (2020).

\bibitem{dpt} M. J. Kronenburg,
A method for fast diagonalization of a 2$\times$2 or 3$\times$3 real symmetric matrix.
arXiv:1306.6291v4 [math.NA] (2015).

\bibitem{dpt2} P. B. Denton, S. J. Parke, T. Tao, and X. Zhang,
Eigenvectors from eigenvalues: A survey of a basic identity in linear algebra.
{\it Bull. Am. Math. Soc.} {\bf 59}(1), 31--58 (2022).

\end{references}
\end{document}